\documentclass{article}
\usepackage{spconf,amsmath,graphicx}
\usepackage{amssymb,color}

\definecolor{nblue}{rgb}{0,0,0.6}
\definecolor{nred}{rgb}{0.8,0,0}
\definecolor{ngreen}{rgb}{0,0.4,0}
\definecolor{norange}{rgb}{1.0,0.5,0.3}

\definecolor{scblue}{rgb}{0,0,0.4}

\newenvironment{Enumerate}{\begin{enumerate}\setlength{\parsep}{0mm}\setlength{\parskip}{0mm}\setlength{\itemsep}{0.5ex}}{\end{enumerate}}
\newenvironment{Itemize}{\begin{itemize}\setlength{\parsep}{0mm}\setlength{\parskip}{0mm}\setlength{\itemsep}{0.5ex}}{\end{itemize}}

\newenvironment{MValue}{\left\{ \begin{array}{ll}}{\end{array} \right.}

\newcommand{\Bf}[1]{{\bf #1}}

\newcommand{\Cal}[1]{{\cal #1}}

\newcommand{\Brace}[1]{\left\{ #1 \right\}}
\newcommand{\Brack}[1]{\left[ #1 \right]}
\newcommand{\Parth}[1]{\left( #1 \right)}

\newcommand{\NTrns}[1]{{#1}^{\mbox{\sf \tiny T}}}

\newcommand{\ParExp}[2]{{#1}^{\mbox{\scriptsize (#2)}}}

\newcommand{\App}{\stackrel{\rm cg}{\approx}}

\newcommand{\Ortho}{{\rm O}(N,\mathbb{R})}

\begin{document}


\title{\vspace{-5mm}LOW-COMPLEXITY TRANSFORM ADJUSTMENTS FOR VIDEO CODING}

\name{\vspace{-4mm}Amir Said \qquad Hilmi E. Egilmez \qquad Yung-Hsuan Chao}
\address{\vspace{-4mm}Qualcomm Technologies, Inc., San Diego, CA, USA}

\maketitle

\begin{abstract}
Recent video codecs with multiple separable transforms can achieve significant coding gains using asymmetric trigonometric transforms (DCTs and DSTs), because they can exploit diverse statistics of residual block signals. However, they add excessive computational and memory complexity on large transforms (32-point and larger), since their practical software and hardware implementations are not as efficient as of the DCT-2. This article introduces a novel technique to design low-complexity approximations of trigonometric transforms. The proposed method uses DCT-2 computations, and applies orthogonal ``adjustments'' to approximate the most important basis vectors of the desired transform. Experimental results on the Versatile Video Coding (VVC) reference software show that the proposed approach significantly reduces the computational complexity, while providing practically identical coding efficiency. 
\end{abstract}


\section{Introduction}

In hybrid video codecs, intra and inter predicted residuals may have diverse statistical properties, which are not fully exploited by symmetric transforms such as the traditional discrete cosine transform (i.e., DCT-2), yet some irregular features can be captured with fixed separable asymmetric transforms including discrete sine transform (DST-7)~\cite{Ye:08:iic,Han:12:jos,Egilmez:16:stb}, or with selections among several separable symmetric and asymmetric transforms by coding a transform index~\cite{Zhao:16:emt}. 
Although asymmetric trigonometric transforms can yield significant coding gains (larger than 1\%), it is also necessary to consider the complexity aspects such as throughput (measured in transform coefficients per second) that can severely limit the maximum frame rate of encoding/decoding. Since throughput decreases with the transform dimension, it is essential to reduce the complexity of large transforms as worst-case measures typically define hardware costs. 

The DCT-2 has been extensively used for image/video coding, because it provides reasonable compression performance and has several low-complexity implementations. Symmetric and regular structure of the DCT-2 leads to efficient hardware implementations (parallel designs) and fast algorithms with low-level software optimizations (e.g., parallel SIMD instructions). Thus, the complexity of DCT-2 can be greatly reduced for critical transform dimensions with fewer arithmetic operations and parallel processing. On the other hand, existing implementations of asymmetric trigonometric transforms are more expensive than the DCT-2 due to their irregular structures~\cite{Britanak:07:dxt}. 

The ITU/MPEG Versatile Video Coding Standard (VVC) at current development stage uses both symmetric and asymmetric transforms DCT-2, DCT-8 and DST-7, where the maximum dimension of DST-7 and DCT-8 is 32. Despite extensive research on efficient computations of trigonometric transforms, most methods do not truly match to the practical video coding requirements and potentially require distinct designs depending on the transform dimension~\cite{Biatek:16:amt}. Specifically, transforms for dimensions up to 16 can be effectively computed with direct matrix-vector products, but it can be prohibitively complex for larger dimensions.

In this paper, we introduce a new approach to design low-complexity approximations of desired transforms by applying a family of efficient transforms closely related to DCT-2 (i.e., DCT-3, DST-2 and DST-3) 
and simple orthogonal transforms called adjustments. The proposed approach is fundamentally different than existing methods~\cite{Britanak:07:dxt}, which directly seek fast implementations of a given transform. On the other hand, our approach uses one of DCT-2, DCT-3, DST-2 and DST-3 as the efficient building block for computation and applies optimized adjustment parameters to better approximate low-frequency basis vectors of a desired transform without degrading the coding performance. Two types of transform adjustment methods are discussed here:\footnote{Both approaches have been proposed for standardization~\cite{Said:18:tas,Said:19:tam}, and extensively tested.}
\begin{Itemize}
\item \emph{Pre-adjustment with sparse band matrices} (Fig.~\ref{fg:SparseAdj}a) first applies adjustment then uses one of DCT-2, DCT-3, DST-2 and DST-3 at the encoder side.
\item \emph{Post-adjustment with block sparse matrices} (Fig.~\ref{fg:SparseAdj}b) first uses one of DCT-2, DCT-3, DST-2 and DST-3 and then applies adjustment at the encoder side.
\end{Itemize}
A non-convex constrained optimization framework is developed to estimate sparse and orthogonal adjustment matrices with a weighted least-squares criterion. Although the proposed method can be used to design any orthogonal transform of interest, here we focus on efficient designs for DST-7 and DCT-8 existing in the latest version of VVC. The proposed adjustment technique has the following main advantages: 
\begin{Itemize}
\item It can greatly reduce the complexity in terms of throughput and memory, since it exploits highly efficient implementations of DCT-2 and sparsity patterns of adjustments (see Fig.~\ref{fg:SparseAdj}). 
\item It can achieve similar coding performance that of a desired transform, since the proposed optimization framework provides very good approximations of the truly important basis vectors.
\end{Itemize}

In the rest of this paper, Section \ref{fg:SparseAdj} describes the proposed optimization framework and discusses the relations between DCT-2, DCT-3, DST-2, and DST-3 by showing that they share the same computational properties. In Section \ref{sc:Results}, the experimental results are presented, and some concluding remarks are drawn in Section \ref{sc:Conclusions}.

\section{Low-complexity transform adjustments}

\begin{figure}
\centering
\includegraphics[scale=0.7]{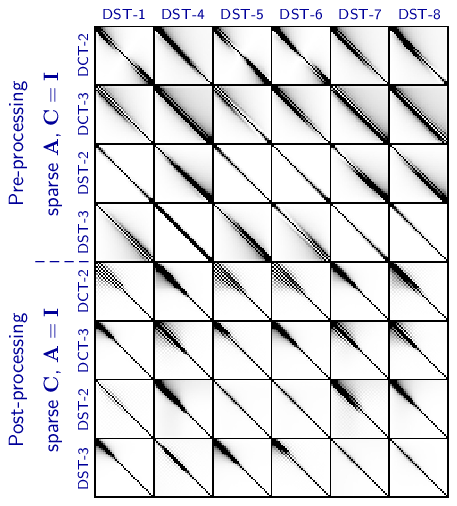}
\caption{\label{fg:MultiAdj}\small Gray-coded representation of sparse adjustment matrices, where larger magnitudes are represented as darker pixels. Along vertical direction are transforms used for computation ($\Bf{B}$), and along horizontal are the desired transforms ($\Bf{D}$).\vspace{-5mm}}
\end{figure}

Representing the group of orthogonal $N\times N$ matrices~\cite{Hall:04:lgl} as $\Ortho$, let $\Bf{D}$ be the desired transform and $\Bf{B}$ be the transform allowing efficient computations. The objective is to find orthogonal transform adjustment matrices $\Bf{A}$ and $\Bf{C}$ such that
\begin{equation}
 \Bf{H} = \Bf{C} \Bf{B} \Bf{A} \App \Bf{D}, \quad \Bf{A} \in \Cal{A}, \Bf{C} \in \Cal{C},
\end{equation}
where $\App$ denotes a form of approximation meant to preserve coding gain, $\Cal{A} \subset \Ortho$ and $\Cal{C} \subset \Ortho$ are predefined subsets of sparse orthogonal matrices that have very low computational complexity.

The following weighted least-squares criterion can be used to achieve similar coding gains that of $\Bf{D}$ by having a matrix approximation with larger weights for low-frequency components.
\begin{equation}
 \label{eq:Optim}
 \min_{\Bf{A} \in \Cal{A}, \Bf{C} \in \Cal{C}} \Brace{ \sum_{i=0}^{N-1} \sum_{j=0}^{N-1} e^{-\alpha i} \Parth{H_{i,j} - D_{i,j}}^2 },
\end{equation}
where $\alpha=\ln(10^2)/N$. The constraint to have matrices belonging to $\Ortho$ (making the optimization problem non-convex) can be satisfied by using a convenient matrix factorization, for example, with sequential sets of parallel Givens rotations~\cite{Said:16:hyg}.

\subsection{Trigonometric transforms related to the DCT-2}

\begin{figure}
\centering
\includegraphics[scale=0.7]{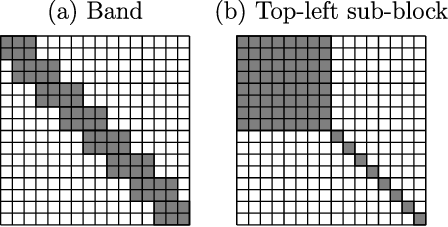}
\caption{\label{fg:SparseAdj}\small Sparsity patterns chosen for DST-7 adjustment matrices: all white-colored elements are equal to zero.\vspace{-5mm}}
\end{figure}

The DCT-2, represented by matrix $\ParExp{\bf T}{C2}$, has several features that are useful for its efficient computation. The DCT-3, DST-2, and DST-3 also have those features because they are respectively related to the DCT-2 according to~\cite{Britanak:07:dxt}
$$
 \ParExp{\bf T}{C3} = \NTrns{\Brack{\ParExp{\bf T}{C2}}}, \; 
 \ParExp{\bf T}{S2} = \Bf{R} \ParExp{\bf T}{C2} \Bf{S}, \;
 \ParExp{\bf T}{S3} = \Bf{S} \NTrns{\Brack{\ParExp{\bf T}{C2}}} \Bf{R},
$$
where the reversal (``flipping'') and interleaved sign changes matrices defined by
\begin{small}
\begin{equation}
 R_{m,n} = 
  \begin{MValue}
   \!\!1, & \!m = N - 1 - n, \\
   \!\!0, & \!\mbox{otherwise,}
  \end{MValue} \;
 S_{m,n} = 
  \begin{MValue}
   \!\!(-1)^m, & \!m = n,\\
   \!\!0, & \!m \neq n,  \end{MValue} \nonumber
\end{equation}
\end{small}
have negligible computational complexity.

\begin{figure*}
\centering
\includegraphics[width=66mm]{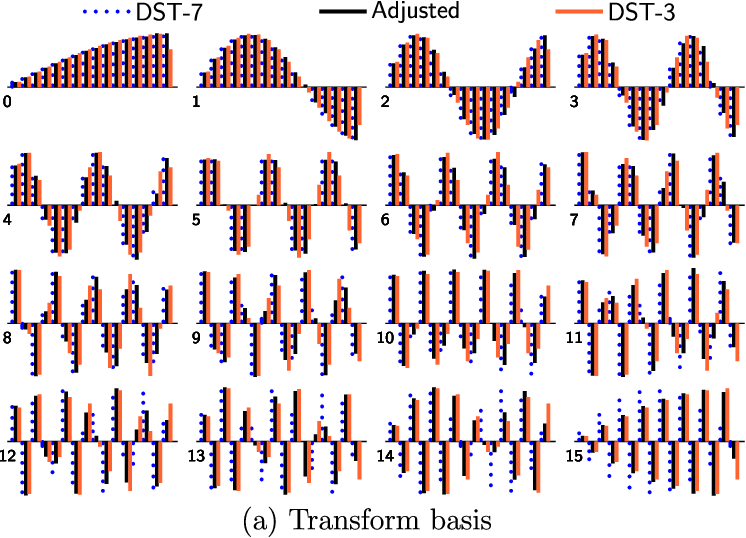} \hspace{6mm} \includegraphics[width=66mm]{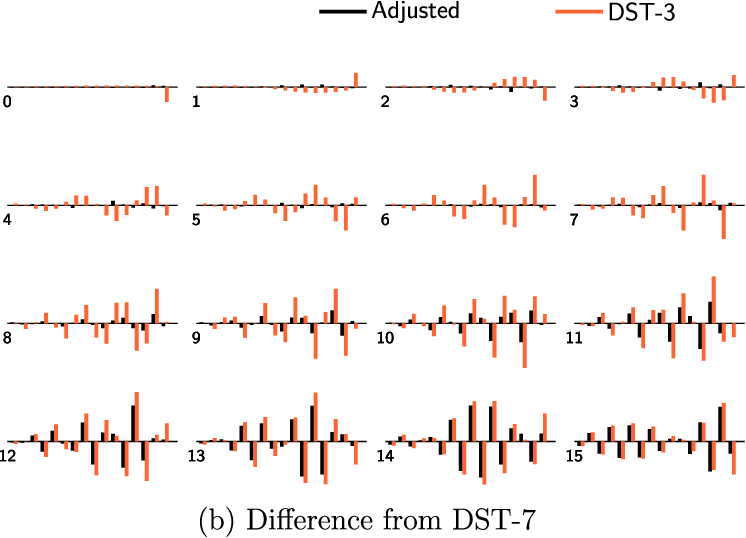}
\caption{\label{fg:BasisComp}\small Example of how the adjustment via band matrix and DST-3 approximates the DST-7, on the transforms of dimension 16. Each subgraph combines the plots of basis functions in increasing frequency.\vspace{-5mm}}
\end{figure*}

While the process of choosing an adjustment transform that simplifies computations may be empiric, there are tools that can be employed for guidance, and only at the final stages it is necessary to use human expertise to select the most practical and effective choice.

For example, if we set $\Bf{C}=\Bf{I}$, we can obtain sparse matrices $\Bf{A}$ by solving an optimization problem like
\begin{equation}
 \min_{\Bf{A} \in \Ortho} \Brace{ \sum_{i=0}^{N-1} \sum_{j=0}^{N-1} \frac{A_{i,j}^2}{0.05 + |A_{i,j}|} + e^{-\alpha i} \Parth{H_{i,j} - D_{i,j}}^2 }. \nonumber
\end{equation}
Fig.~\ref{fg:MultiAdj} show examples of the results of such optimizations when the DCT-2, DCT-3, DST-2, and DST-3 are used as matrix $\Bf{B}$, and the remaining DSTs as matrix $\Bf{D}$, and when either $\Bf{A}=\Bf{I}$ or $\Bf{C}=\Bf{I}$ (results for other DCTs are not shown because they are identical, but permuted versions).

The resulting sparsity indicates how well the approximation is expected to work. While various degrees of sparsity are obtained, note that in each column there is at least one reasonably sparse case.

\subsection{Transform adjustments for the DST-7}

The approach described in the previous sections has been successfully  used to replace several trigonometric transforms~\cite{Said:18:tas}. In this article we show results for the VVC asymmetric transforms, i.e., the DST-7 and DCT-8, represented by matrices $\ParExp{\bf T}{S7}$ and $\ParExp{\bf T}{C8}$. We can focus only on the DST-7 because the DCT-8 is simply its ``flipped'' version, i.e.,
\begin{equation}
 \label{eq:DCT8flip}
  \ParExp{\bf T}{C8} = \Bf{S} \ParExp{\bf T}{S7} \Bf{R}.
\end{equation}

Using the insight obtained from Fig.~\ref{fg:MultiAdj}, the types of sparse adjustment matrices chosen for approximating the DST-7, because they can be efficiently computed, are as shown in Fig.~\ref{fg:SparseAdj}:
\begin{Enumerate}
 \item a band matrix $\Bf{A}$ before the DST-3;
 \item a top-left sub-block matrix $\Bf{C}$ after the DCT-2.
\end{Enumerate}

Since the first approach has computations similar to bi-orthogonal filters, band matrices with at most $K$ non-zero row elements are referred as $K$-tap adjustments.

The second approach is similar to secondary transforms proposed by other authors~\cite{Alshina:11:rot,Saxena:12:sti,Saxena:13:lls}, but it should be noted that, while in a $W \times H$ block a secondary separable transforms may be applied horizontally and vertically to a $B \times B$ sub-block, here the transform adjustments are applied horizontally and vertically to $W \times B$ and $B \times H$ blocks. Our experimental result show that this difference is important.

Note that we tested only one or the other adjustment at a time ($\Bf{A}$ or $\Bf{C}$), and not both simultaneously. In each case what is obtained is similar to the example in Fig.~\ref{fg:BasisComp}, where the adjustment is done with a 6-tap band matrix. Note that the basis functions with lowest frequencies are remarkably well approximated, while the others, which are much less important in defining coding gains, are not.

\subsection{Simultaneous transform adjustments for the DCT-8 and DST-7}\label{sc:Simult}

When the sub-block matrix of Fig.~\ref{fg:SparseAdj}b is used for adjustment matrix $\Bf{C}$ after the DCT-2, we can exploit the fact that
\begin{equation}
  \Bf{S} \ParExp{\bf T}{C2} = \ParExp{\bf T}{C2} \Bf{R},
\end{equation}
to conclude that if $\ParExp{\bf C}{S7}$ is the adjustment matrix for DST-7, then from eq.~(\ref{eq:DCT8flip})
\begin{equation}
 \label{eq:AdjSymm}
 \ParExp{\bf C}{C8} = \Bf{S} \ParExp{\bf C}{S7} \Bf{S}.
\end{equation}
This means that the magnitudes of the elements of those matrices are the same, but the signs are different according to a chessboard pattern.

Since the encoder needs to evaluate, for a given block of residuals, the two-dimensional DCT-2 and all combinations of the DCT-8 and DST-7 horizontally and vertically, it may need to re-compute five transforms for each block to choose the best.

The property defined by eq.~(\ref{eq:AdjSymm}) enables the simultaneous computation of DCT-8 and DST-7 adjustments with same number of arithmetic operations as needed for one adjustment. In addition, since this type of adjustment is applied to only the first 8 lines in a block, for dimension 32 and larger, less than 50\% of the transform coefficients are changed for all combinations, and thus do not need to be recomputed. This efficient encoder implementation has also been tested in the VVC codec, and more details are available in ref.~\cite{Said:19:efc}.

\section{Experimental results}\label{sc:Results}

\begin{table}
\centering
\caption{\small \label{tb:Through}Throughput measurements of transform adjustments when used in the VTM-3.0 software, compared to full matrix computation.}
\begin{small}
\begin{tabular}{|l||c|c||c|c|} \hline
 Adjustment & \multicolumn{4}{|c|}{Transform throughput ratios} \\ \cline{2-5}
 type & \multicolumn{2}{|c||}{Forward} & \multicolumn{2}{c|}{Inverse} \\ \cline{2-5}
                   &  32-pt  &  64-pt  &  32-pt  &  64-pt  \\ \hline \hline
Band, 4-taps       & 1.20 & 0.85 & 2.62 & 2.24 \\ \hline
Band, 6-taps       & 1.14 & 0.79 & 2.44 & 2.08 \\ \hline
$8\times 8$ matrix & 1.38 & 1.60 & 5.55 & 5.74 \\ \hline
Multiple           & 2.67 & 5.56 & 5.55 & 5.74 \\ \hline
\end{tabular}
\end{small}
\end{table}

\begin{table*}
\centering
\caption{\small \label{tb:VTM_results} All-intra coding gains of using DCT-2, DST-7, DCT-8 over using only DCT-2 in BD-rate (\%).}
\begin{small}
\begin{tabular}{|c||c|c|c|c|c|c|c|c|} \hline
Transform  & \multicolumn{8}{|c|}{Classes of video sequences} \\ \cline{2-9} 
dimensions &    A1  &   A2   &    B   &    C   &    D   &    E   &    F   &  CTC avrg. \\ \hline 
 CTC        &   3.05 &   2.67 &   2.76 &   2.04 &   1.91 &   3.34 &   1.69 &   2.73  \\ \hline
CTC + 64-pt &   3.22 &   2.74 &   2.78 &   2.04 &   1.91 &   3.35 &   1.69 &   2.78  \\ \hline 
\end{tabular}
\end{small}
\end{table*}

\begin{table*}
\centering
\caption{\small \label{tb:GainResults} All-intra coding gain results of proposed method over VTM-3.0 under CTC in BD-rate (\%).\vspace{4mm}}
\begin{small}
\begin{tabular}{|c||c|c|c|c|c|c|c|c|} \hline
Applied    & \multicolumn{8}{|c|}{Classes of video sequences} \\ \cline{2-9} 
dimensions &    A1  &   A2   &    B   &    C   &    D   &    E   &    F   &  CTC avrg. \\ \hline \hline
 \multicolumn{9}{|c|}{4-tap band matrix adjustment (BD-rate gain difference) } \\ \hline
32         & --0.05 & --0.10 & --0.03 & --0.01 & --0.02 & --0.03 & --0.01 & --0.04 \\ \hline
64         &  +0.17 &  +0.07 &  +0.03 &  +0.01 &   0.00 &  +0.02 &  +0.01 &  +0.05 \\ \hline
32, 64     &  +0.15 & --0.03 & --0.00 & --0.01 & --0.02 & --0.01 &  0.00 &   +0.02 \\ \hline \hline
 \multicolumn{9}{|c|}{6-tap band matrix adjustment (BD-rate gain difference) } \\ \hline
32         & --0.01 & --0.04 & --0.02 & --0.01 & --0.03 & --0.02 & --0.01 & --0.02 \\ \hline
64         &  +0.17 &  +0.07 &  +0.03 &  +0.01 &   0.00 &  +0.03 &  +0.01 &  +0.06 \\ \hline
32, 64     &  +0.15 &  +0.04 &  +0.02 &   0.00 & --0.02 &  +0.02 &   0.00 &  +0.04 \\ \hline \hline
 \multicolumn{9}{|c|}{$8\times 8$ sub-matrix adjustment (BD-rate gain difference) } \\ \hline
32         &  +0.14 & --0.08 & --0.04 & --0.01 & --0.02 &   0.00 & --0.01 &   0.00 \\ \hline
64         &  +0.33 &  +0.07 &  +0.04 &  +0.02 &   0.00 &   0.02 &   0.00 &  +0.08 \\ \hline
32, 64     &  +0.45 & --0.02 &  +0.01 & --0.02 & --0.02 &  +0.02 &   0.00 &  +0.07 \\ \hline
\end{tabular}
\end{small}
\end{table*}

The parameters for band matrix adjustments used for the experimental results shown in this section are listed in~\cite{Said:18:tas}, while the parameters for sub-block matrices are in~\cite{Said:19:tam}.\footnote{JVET documents are available at http://phenix.it-sudparis.eu/jvet/.}

In the latest VVC reference software, VTM-3.0~\cite{Chen:18:vtm}, the maximum DST-7 and DCT-8 size is restricted to 32, and it is 64 for DCT-2. As shown in Table~\ref{tb:VTM_results}, the transform scheme in VTM-3.0 under common test conditions (CTC)~\cite{Bossen:18:ctc} provides 2.73\% coding gain on average in BD-rate \cite{Bjontegaard:01:cap} as compared to using the DCT-2 only. Although 64-point DST-7 and DCT-8 would provide about 0.2\% additional coding gains for class A1 and A2 sequences with ultra high-definition (UHD) resolution (see ``CTC + 64" in Table \ref{tb:VTM_results}), those are not part of VTM-3.0 due to their excessive complexity and memory requirements. In our experiments, we design 32 and 64-point DST-7 and DCT-8 with the proposed adjustment technique, so that 64-point DST-7 and DCT-8 are enabled without increasing the worst-case complexity (in terms of number of multiplications per coefficient), and the complexity of 32-point DST-7 and DCT-8 are considerably reduced~\cite{Said:19:tam}. 

Table~\ref{tb:GainResults} presents all-intra coding results of the proposed adjustment methods over the VTM-3.0 under CTC\footnote{Experimental results for random-access and low-delay configurations, and also very high bit-rates, are presented in~\cite{Said:18:tas,Said:19:tam}.}. The results show that proposed transform adjustment methods provide very good approximations for 32 and 64-point DST-7 and DCT-8, so that they have very little or no effect on compression performance (see the rows labeled as ``32" and ``64" in Table~\ref{tb:GainResults}). While the adjustment using 6-tap band matrix provide slightly better approximations on average as compared to the other two cases, adjustments with 4-tap band and 8x8 matrix have lower complexity.

Table~\ref{tb:Through} shows the ratio of transform throughputs, compared to the direct matrix-vector multiplication, using a benchmarking method described in~\cite{Said:19:stt}. The line ``multiple'' refers to the simultaneous computation of the DCT-8 and DST-7 at the encoder, as defined in Section~\ref{sc:Simult}. Those throughput measures demonstrate the significant benefit of adjustment with 8x8 matrix, which is more than 5 times faster than the full matrix multiplication-based implementations in VTM-3.0 and is more than 2 times faster than adjustments with 4-tap and 6-tap band matrices. 

\section{Conclusions} \label{sc:Conclusions}

This paper proposes a new method method for low-complexity approximations of desired transforms and introduces two variants of adjustment based on band and sub-block transform matrices. The proposed methods are used to reduce the complexity of large DST-7 and DCT-8 and tested on a real codec (VTM-3.0). Inspection of the experimental results and our complexity analysis 
lead us to the following conclusions:
\begin{Itemize}
\item Both variants of the proposed approach provide a very good approximation of DST-7 and DCT-8 with negligible coding performance differences on VTM-3.0.
\item As compared to using band matrices, the 8x8 proposed subblock-based design provides a better trade-off in terms of performance and complexity. 
\item With the 8x8 subblock adjustment, the worst-case complexity of VTM-3.0 is reduced to 36 multiplications per-coefficient from 64 and the memory reduced is reduced by 1 kilobytes. 
\item Enabling 64-point DST-7 and DCT-8 has no impact on the worst-case complexity and provides up to 0.45\% coding gains for ultra high definition (UHD) videos.  
\end{Itemize}


\clearpage

\begin{small}
\bibliographystyle{IEEEbib}
\bibliography{mmsp}
\end{small}

\end{document}